\documentclass[9pt,conference]{IEEEtran}
\usepackage{amssymb,amsthm,amsmath,array}
\usepackage{graphicx}
\usepackage[caption=false,font=footnotesize]{subfig}
\usepackage{xspace}
\usepackage[sort&compress, numbers]{natbib}
\usepackage{stmaryrd}
\usepackage{xcolor}
\usepackage{mathtools}
\usepackage{float}
\usepackage{textcomp}

\begin{document}
\title{High-resolution single-pixel imaging in real time with iterative or deep learning-based reconstruction enhancement}
\author{\IEEEauthorblockN{
        Anna Pastuszczak\IEEEauthorrefmark{1}, 
        Rafał Stojek\IEEEauthorrefmark{1,2},  
        Piotr Wróbel\IEEEauthorrefmark{1},
        Magdalena Cwojdzińska\IEEEauthorrefmark{1},
        Kacper Sobczak\IEEEauthorrefmark{1}, and
        Rafał Kotyński\IEEEauthorrefmark{1}
    }
    \IEEEauthorblockA{
        \IEEEauthorrefmark{1} Faculty of~Physics, University of~Warsaw, Pasteura 5, 02-093 Warsaw, Poland.\\
        \IEEEauthorrefmark{2} VIGO Photonics, Poznańska 129/133, 05-850 Ożarów Mazowiecki, Poland.}
}
\maketitle
\begin{abstract}
We introduce a compressive single-pixel imaging (SPI) framework for high-resolution image capture in fractions of a second. This framework combines a dedicated sampling strategy with a tailored reconstruction method to enable high-quality imaging of spatially sparse scenes at the native 1024×768 resolution of a digital micromirror device (DMD). The reconstruction process consists of two phases: first, the measured data is processed using the generalized inverse of the measurement matrix for quick image recovery. Then, the spatial sparsity of the scene is leveraged to enhance reconstruction in dense areas, using either an iterative method or a neural network-based approach. With a compression ratio of 0.41\% and an image acquisition rate of 6.8 Hz at 22 kHz DMD operation, this framework supports real-time, high-resolution dynamic imaging with the reconstruction that matches the acquisition rate on a mid-tier desktop GPU.
\end{abstract}

\section{Introduction}
Single-pixel imaging (SPI)~\cite{Duarte_2008, Gibson2020} uses high-speed spatial light modulation to indirectly measure the spatial resolution of a scene with a single photodetector. Although SPI gains attention for various applications, such as imaging across different wavelength ranges~\cite{Olbinado:21, 2019OExpr.27.4562D, Stantchev2020, Zanotto:OEA2020}, spectral imaging~\cite{Klein:23, Ghezzi:21}, imaging through complex media~\cite{Ipus:24}, 3D imaging~\cite{Qian:19, Salvador-Balaguer2018}, etc., the typical resolution of SPI systems falls far below that of classical imaging cameras and rarely exceeds $256 \times 256$ pixels. In contrast, digital micromirror devices (DMDs), which are commonly used in SPI due to their high operational bandwidth (typically slightly above 20 kHz), offer resolutions of $1024 \times 768$ or higher.
The main limitation for high-resolution SPI systems (in terms of the number of pixels) is the trade-off between modulation bandwidth, the compression ratio, and the computational complexity of image reconstruction. Imaging at the native resolution of the DMD either requires severe compression, compromising image quality, or results in prohibitively long acquisition times, making it impossible to capture dynamic scenes. For instance, in recent work~\cite{Li:ol2023}, capturing a single $1024 \times 768$ high-quality image using the Fourier SPI technique takes as long as 1180~s.

To confront these challenges, we have recently proposed an advanced SPI framework that enables high-resolution imaging of spatially sparse scenes in only fractions of a second~\cite{Stojek:sensors2024}. This framework involves sensing at an extreme compression ratio of 0.41\% using carefully designed binary sampling patterns, similar but not identical to those in our previous work~\cite{Stojek:OE-30-20009-22}, and a complementary image reconstruction algorithm that allows for the quick identification of empty sectors in the image and enhancement of reconstruction quality in the non-empty areas. The reconstruction process is divided into two stages. First, an initial low-quality image is obtained using the Fourier-domain regularized inverse of the sensing matrix, which enables fast image reconstruction with a single matrix-vector product~\cite{Czajkowski_2018OE}. Next, we employ either a dedicated iterative algorithm or a deep learning approach to leverage the spatial sparsity of the image and refine the reconstruction quality. Deep learning has been successfully applied to SPI in previous works to improve image reconstruction~\cite{Higham2018, Wu:20, Ni:OL2021}.

\section{Methodology}
In this section, we recall and discuss the high-resolution SPI framework originally proposed in~\cite{Stojek:sensors2024}.

The sampling method in this framework is based on highly compressive differential binary sampling patterns derived from multiple partitions of the image space. Each partition, referred to as an image map, is a mapping between image pixels and a subset of integer values, which divides the image plane into regions where all pixels share the same value. A region may contain more than one group of adjacent pixels or even isolated pixels. The image maps are generated through uniform quantization of the phase of multiple realizations of correlated complex Gaussian noise with varying characteristic sizes. However, the choice of specific image maps in this framework is arbitrary, and other partitions of the image space, including non-uniform ones, are also possible~\cite{Pastuszczak:Cosera2024}.

The image maps are translated into binary sampling patterns using an auxiliary binary matrix that functions as a look-up table. Each sampling pattern consists of a linear combination of approximately half of the regions from an image map, selected with the help of the look-up table. This results in each sampling pattern having approximately 50\% of its pixels set to 1 (or 50\% of the DMD mirrors in the 'on' state). Designing all the sampling patterns to have a similar number of pixels in the 'on' state greatly improves the SNR in the SPI measurement by enhancing the entropy of the measured signals, making better use of the detector's measurement range, and reducing quantization errors induced by the DAQ.

The main advantage of the proposed sampling method is that it not only enables the extraction of spatial information about the scene in an extremely compressive SPI measurement, but also allows us to easily identify empty and non-empty regions in the image. This property arises from the fact that the proposed sampling method is equivalent to individually sampling the image using all regions of the image maps separately, which allows for the accurate measurement of mean pixel values in each region of every map (in a noiseless measurement scenario). We have developed an algorithm that identifies all regions of the image with mean pixel values close to 0. These regions are regarded as empty, and all pixels within them are explicitly set to zero. The algorithm then iteratively rescales the pixel values in the remaining non-empty regions to restore their correct mean values. This step significantly improves the quality of the reconstruction for spatially sparse images, nearing the level of detail provided by the full resolution of the DMD. Typically, 1 to 5 iterations are sufficient for optimal reconstruction improvement.

In an alternative approach, we investigated replacing the iterative algorithm with a post-processing step based on a neural network (NN) inspired by a U-net architecture. The NN was trained using a loss function that combines two components: (1) the mean squared error (MSE) of the pixel values as compared to the ground truth — to eliminate reconstruction artifacts caused by measurement noise and compression, and (2) the MSE of the mean values of pixels in all regions derived from the image maps — to promote sparsity and utilize the unique properties of the proposed sampling method. The training dataset is composed of two types of spatially sparse images: (1) binary, high-resolution line art images, and (2) subsets of randomly placed and scaled handwritten digits drawn from the MNIST dataset \cite{MNIST2012}. 
The NN-based approach is significantly faster, with computation time nearly six times shorter than a single iteration of the original algorithm. Moreover, with GPU-accelerated computation, both the NN post-processing and a single iteration of the iterative algorithm support real-time image acquisition and reconstruction at 6.8 Hz with a resolution of 1024 × 768, without limiting the DMD acquisition rate.

\section{Results} 

\begin{figure}[tbp]
    \centering
    \includegraphics[width=1\linewidth]{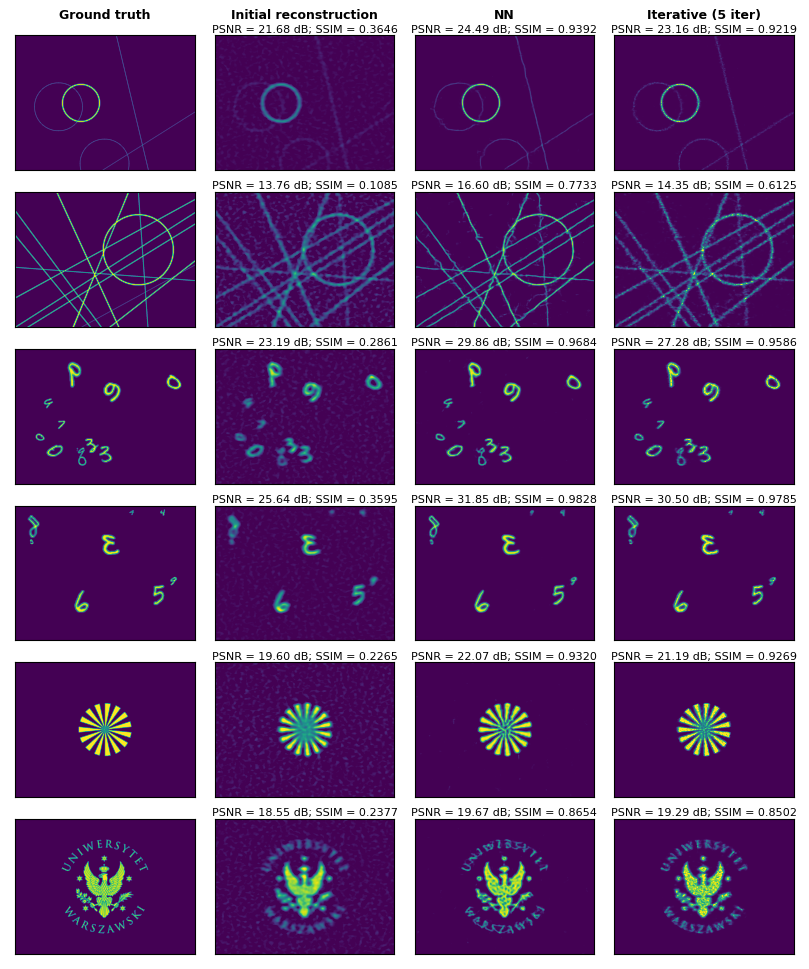}
    \caption{Comparison of image reconstructions obtained using either the neural network (NN) or the iterative algorithm (with 5 iterations) as the reconstruction enhancement method. Six spatially sparse images with varying levels of complexity were used to simulate a compressive SPI measurement without measurement noise. The reconstruction quality for both methods, as well as for the initial reconstruction (included for comparison), was evaluated using PSNR and SSIM metrics.}
    \label{fig:rec_comparison_noiseless}
\end{figure}

\begin{figure}[tbp]
    \centering
    \includegraphics[width=1\linewidth]{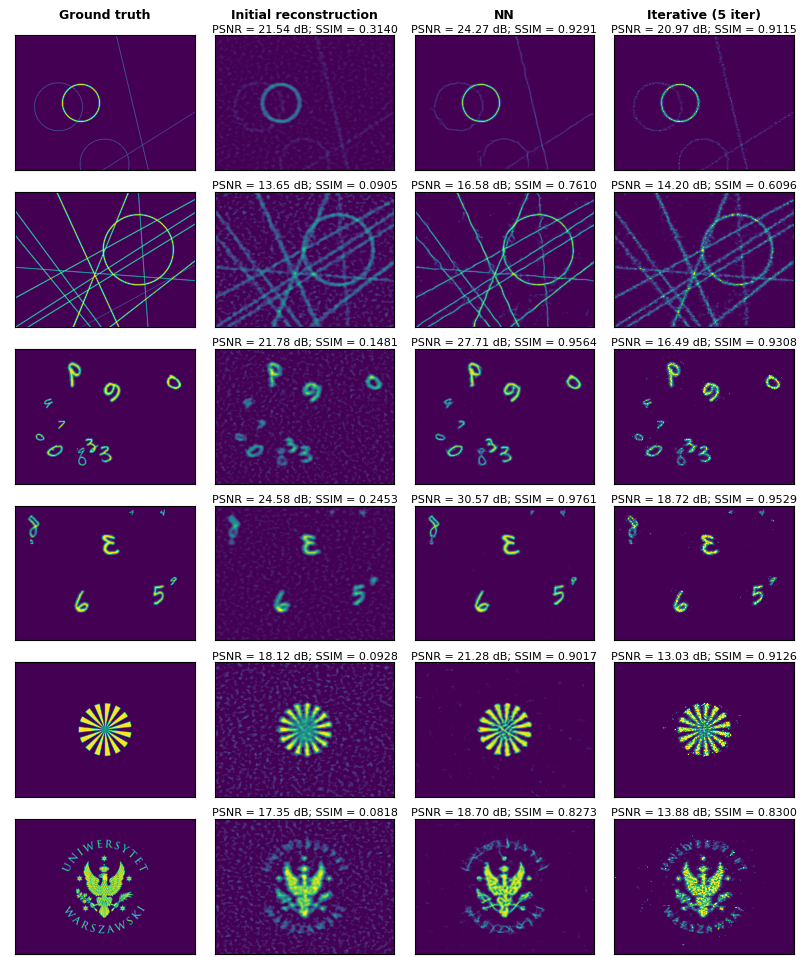}
    \caption{Comparison of image reconstructions obtained using either the neural network (NN) or the iterative algorithm (with 5 iterations) as the reconstruction enhancement method. The same six spatially sparse images as shown in Fig.~\ref{fig:rec_comparison_noiseless} were used to simulate a compressive SPI measurement in the presence of additive Gaussian detector noise with an SNR of 46 dB.}
    \label{fig:rec_comparison}
\end{figure}

In Figs.\ref{fig:rec_comparison_noiseless} and \ref{fig:rec_comparison}, we compare the performance of the two proposed image reconstruction enhancement methods both qualitatively and quantitatively using two common metrics: peak signal-to-noise ratio (PSNR) and structural similarity index (SSIM). Six spatially sparse images of varying complexity were selected for the comparison: the top four images are from the NN testing dataset and are similar (but not identical) to the training data, while the bottom two images (a vignetted Siemens star resolution chart and the logo of the University of Warsaw) differ significantly from the training data. Fig.\ref{fig:rec_comparison_noiseless} shows reconstructions from a simulated noiseless compressive SPI measurement, whereas Fig.~\ref{fig:rec_comparison} includes realistic additive detector noise comparable to that in our experimental setup.
The comparison shows that both methods significantly improve the resolution and visual quality of the image reconstructions compared to the initial reconstruction stage. The iterative method primarily relies on accurately measured mean pixel values across all image regions derived from the image maps, making it more susceptible to measurement noise. In contrast, the NN generally achieves higher PSNR and SSIM scores but may introduce significant reconstruction artifacts when processing images that differ from the training data. The iterative algorithm, however, is more versatile and can handle imaging scenes with arbitrary content.

\section{Conclusion}
We have introduced a compressive SPI framework for high-quality imaging of spatially sparse scenes at the native 1024×768 resolution of a DMD, achieving an image acquisition and reconstruction rate of 6.8 Hz. The framework features a differential, binary, non-adaptive sampling strategy based on multiple subdivisions of the image space, paired with a two-stage reconstruction algorithm that exploits object sparsity and sampling properties to identify empty regions of the image and refine non-empty areas. Reconstruction enhancement is achieved using either iterative or NN-based processing as part of the proposed algorithm. While both methods significantly improve reconstruction quality, the NN approach is computationally faster, whereas the iterative method is more versatile, handling both varying measurement conditions and images with diverse content.

\bibliographystyle{IEEEtran}
\bibliography{references}
\end{document}